# Psychometric properties of an instrument to investigate the motivation of visitors to a science museum: The combination of methods

Rosana de Fátima Martinhão[1], Kenia Naara Parra*[1], Daniela Maria Lemos Barbato[1,2]

and Ana Cláudia Kasseboehmer[1]

[1]Instituto de Química de São Carlos, Universidade de São Paulo, Brazil

[2]Instituto SEB de Educação, Ribeirão Preto, Brazil

The visit to a science museum may be manifested through complex and dynamic motivations which, according to the literature, are under-investigated in a Brazilian context. In the present study, an instrument originally developed by Delgado in 2008 (http://hdl.handle.net/10773/1623) has been modified and applied to 202 visitors up to 15 years to the Science Museum "Professor Mário Tolentino" in São Carlos, Brazil, in order to investigate motivation for visiting the institute. Combined application of Exploratory Factor Analysis and the Information Bottleneck method revealed that 17 out of the 20 initial items in the questionnaire aligned with three dimensions of motivation. The main motivation was learning desire, while entertainment and interaction motivations were significantly less important. The study provided relevant evidence regarding the motivations of visitors, and this information will be valuable in improving the activities of the museum. The implications of our findings for future research are discussed.

______________________________________

*Corresponding author; e-mail: keniaparra@usp.br

# INTRODUCTION

Visitors utilize the experience of a museum visit in order to confirm their current understanding, to acquire new knowledge, to achieve self-fulfillment, to participate in museum activities, to interact with others, to learn about the emblematic destinations of a specific country or region, to have fun or a different type of day out with their children, or simply just to "kill time" (Delgado, 2008; Ji et al., 2014; Packer & Ballantyne, 2002). The determination of exactly what visitors want from their museum visit, and how satisfied they are with the experience, can be achieved by establishing the motivation behind the visit. It is of interest, therefore, to consider the type of instruments that have been applied for the purpose of understanding the nature of motivation within a setting such as a science museum.

## Motivation: a Construct to be Evaluated

Visitor research studies reflect an increasing interest in understanding the motivation and behavior of visitors to informal educational exhibitions (Zwinkels, Oudegeest, & Laterveer, 2009). Both qualitative and quantitative approaches have been employed in order to augment the effectiveness and agility of the educational and cultural goals of these institutions from psychological, educational, social and communicational perspectives (Studart, 2005). According to this author, the results of visitor motivation research have engendered increases in audience extent mediated by changes in physical infrastructure as well as by the incorporation of more leisure and differentiated services.

It is, therefore, important to establish a complex sociological and psychological construct relating to a theoretical concept about the nature of human behavior. Psychological constructs such as intelligence, ambition and motivation are referred to as latent variables because they are not directly observable but can be measured by means of items or statements related to the construct that serve as empirical indicators of how the construct is conceptualized (Glynn, Taasoobshirazi, & Brickman, 2009).

Conceptualizations of a construct such as motivation for visiting a science museum are important because they influence the actions of visitors and appear to reflect their experience of the visit. Moreover, studies aimed at understanding the motivation for visiting a science museum could also evince new and improved ways to attract visitors.

Falk and Dierking (1992, 2000) proposed an "interactive experience model" to interpret both the motivation to visit museums and the museum experience based on the interaction of socio-cultural, physical, personal and temporal contexts. Through these contexts it is possible to assess learning in museums and the relationships that can occur before, during and after the visit. More recently, Falk (2006) identified the existence of identity-related roles and needs of visitors that influence how they perceive the museum and what role it should play in the lives. The author proposed clustering all of the various visitor motivations into five distinct identity-related categories such as explorers, facilitators, professional/hobbyists, experience seekers and recharges.

With the aim of identifying people who are poorly motivated for visiting museums and the reasons for such depleted motivation, Delgado (2008) studied the constraints that inhibit or block visits to the "Pavilion of Knowledge" in Lisbon, Portugal. The author found structural, personal and interpersonal constraints that discourage the visitor and suggested five categories of motivation for visiting science museum, namely interaction and entertainment, discovery news, learning, self-realization and participation in museum activities. These findings served as a basis for changes in the studied science center and also provided a tool for understanding the motivations for visiting museums that could be evaluated worldwide.

More recently, Ji et al. (2014) reviewed the motivational components that influence visits to museums and aquariums in Beijing, China and in Vancouver, Canada. Based on the expectations and requirements that participants gave for visiting these institutes, the authors suggested that motivation tended to cluster around five dominant components, namely education, entertainment, personal interest, social interaction, and practical issues (such as tourist attraction).

**Under-representation of Motivational Studies in the Brazilian context**

Various studies of the motivation for visiting science centers have been undertaken in the cultural contexts of North America, Europe and, more recently, East Asia. The results indicate that different contexts and types of informal settings reflect different motivations. For example, studies in a non-Western context revealed that Chinese families valued the opportunities for childhood education more highly than the possibility of social interaction with others, while in a Western context, Chinese families considered social motivation to be of greater importance (Ji et al., 2014). In contrast, research developed by Lin (2006) revealed that, from a Taiwan perspective, the barrier to visiting museums was the strong association of such institutes with education and learning rather than with entertainment and exploration.

In Brazil, the Federal Government has made a strong financial commitment, through the Ministry of Science, Technology and Innovation (MCTI), to constructing and refurbishing science museums. According to the 2015 edition of the *Guide of Centers and Science Museums of Brazil* (Associação Brasileira de Centros e Museus de Ciência, 2015), there are 268 science centers registered across the country. However, these centers are poorly distributed across the country with 58% concentrated in the southeast region. Moreover, the centers remain poorly utilized by local populations and the MCTI states that 96% of Brazilian people have never visited a museum.

Information concerning the attendance at science museums in Brazil is sparse or only available in the form of local speculative studies that are not available internationally. With the aim of contributing to an understanding of the Brazilian audience, we tested the validity of the instrument for evaluating the motivation for a museum visit developed by Delgado (2008), and surveyed the motivation for visiting the Science Museum "Professor Mário Tolentino" in São Carlos, SP, Brazil. This study constitutes part of a larger research project entitled "Development of a Chemistry Gallery".

**METHODS**

**Study Site**

The Science Museum "Professor Mário Tolentino" located in the city of São Carlos, SP, Brazil, (Figure 1) has free admission and visits can be made spontaneously or scheduled. The museum receives an average of 170 people a day from the local community and from cities near to São Carlos.

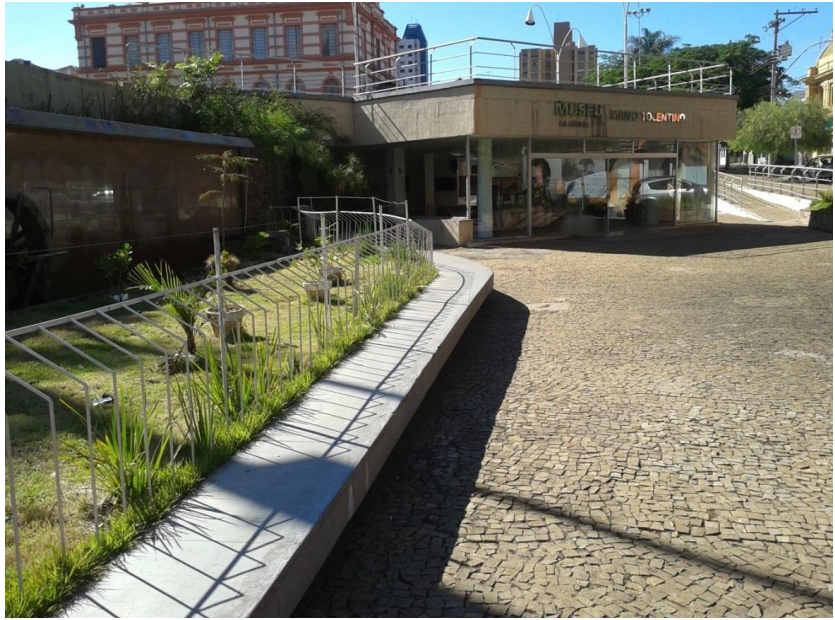

**Figure 1.** The entrance to the Science Museum "Professor Mário Tolentino" in São Carlos, SP, Brazil.

Although the museum covers an area of 2,200 m$^2$, it currently houses permanent exhibitions only in fields of physics and paleontology. Substantial space is available for expansion and development of other exhibitions dealing with, for example, chemistry, astronomy or biology. A deeper understanding of the motivation behind visits to the museum will contribute to the development of an interactive space that could attract more visitors.

**Instrument Design**

In the course of the present study we employed a two-part survey. In part A, we gathered personal information with nine questions about the level of schooling, occupation, type of transport used to get to the museum and other complementary data. Part B of the survey related to the motives for visiting the museum and the sources of motivation were adapted from a scale based on the motivational components employed in the measurement of this construct in a science center in Lisbon, Portugal (Delgado,

2008). The questionnaire, with five point agreement-disagreement Likert-type scale, initially contained 20 items divided between five hypothetical dimensions but distributed randomly. As shown in Appendix 1, the five subscales measured the *Leisure and entertainment motivation* (five items; sample item "Being with people"), *Discovery news motivation* (four items; sample item: "Satisfying my curiosity about science"), *Learning motivation* (five items; sample item: "Increasing my knowledge and that of my family/friends"), *Interaction motivation* (four items; sample item: "Interacting with people") and *Participation in museum activities motivation* (two items; sample item: "Participating in museum activities").

The wording and types of items and dimensions were tested and agreed upon by education researchers in Brazilian science museums. However, staff at the study museum vetted four specific questions in the complementary data section and recommended that information relating to age and home address should be obtained from the museum visitor's book rather than from the questionnaire.

Although all of the items presented in this paper are in shown English, the questionnaire was developed in Portuguese. In order to check the quality of our translation, all items were back-translated from English and the resulting Portuguese version was faithful to the original wording.

**Participants and Data Collection**

The study was conducted at different times of the year including school vacation (July 2014) and schooling periods (September, October and November 2014). The reason for choosing these periods was that the museum draws in a large number of people, including families from the local community and surrounding cities, who may not usually visit a science museum.

Data were collected on-site and in written form immediately before visitors entered the exhibition. Visitors were made aware that participation in the survey was voluntary and that no identifying information would be divulged to third parties. To allow for maximum data collection, the questionnaire was self-administered by the participants.

A total of 202 usable surveys were collected from male and female visitors of minimum age 15 years, the majority of whom ($n = 105$; 51.98%) were visiting the museum for the first time. Among the respondents, 154 (76.24 %) were aged between 15 and 24 years, 23 (11.39 %) between 25 and 34 years, 16 (7.92 %) between 35 and 44 years, 6 (2.97 %) between 45 and 54 years, 2 (0.99 %) between 55 and 64 years, and 1 (0.50 %) was aged 65 years or more. Females ($n = 125$; 61.88 %) and students ($n = 168$; 83.17 %) constituted the largest groups of respondents.

**Statistical Methods**

In the case of psychometric instruments, the sample variability may cause different factorial solutions, even for the same questionnaire. Thus, evaluating the factorial structure of a psychometric instrument is important to establish a complex sociological and psychological construct (Massidda, Carta, & Altoè, 2016; Bjørnebekk, 2009; Tze et al., 2013).

The factor structure analysis of the psychometric instrument proposed by Delgado (2008) was developed through two separate types of statistical treatment, namely exploratory factor analysis (EFA) to investigate latent variables (factors) and information bottleneck (IB) analysis as a comparative method to investigate the grouping structure of the items.

*Exploratory Factor Analysis*

Since the questionnaire had not been validated, we applied EFA rather than confirmatory analysis (Glynn, Taasoobshirazi, & Brickman, 2009) in order to establish the existence of a small number of factors based on a larger number of latent variables, to eliminate mutual correlation among dependent variables and to obtain a highly reliable research tool (Costello & Osborne, 2005; Williams, Onsman, & Brown, 2010).

The first step of our analysis was to investigate the reliability of the questionnaire by calculating Cronbach's alpha coefficient. Values greater than 0.7 are considered acceptable (Robinson, Shaver, & Wrightsman, 1991). Before applying EAF, we verified correlation among the data using Bartlett's sphericity test in order to reject the null hypothesis of identity matrix, and the Kaiser-Meyer-Olkin (KMO) statistic to calculate the level of correlation among the variables. Values of the KMO statistic can range between 0 and 1, and values greater than 0.8 are considered acceptable (Tabachnick & Fidell, 2007).

Factor extraction was carried out by applying principal component analysis to calculate eigenvalues of the data correlation matrix. These eigenvalues are associated with the factors and can explain the variance for each factor. Various criteria have been suggested for choosing the number of eigenvalues to be retained. Hair et al. (2006) state that more than one criterion should be applied taking into account theoretical foundations and experimental evidence. Once the number of factors had been established, we investigated factor structure by analyzing the factor loading of each item (observable variable), the value of which depends on sample size (Hair et al., 2006). It is sometimes necessary to rotate the factor matrix in order to understand the contribution of an item to each factor. If the structure remains even after rotation, the item should be rejected. At the end of this process the observable variables (items) were reduced to latent variables or factors, which were named according to the items grouped therein.

*Information Bottleneck Method*

Since the questionnaire had not been validated, a comparative analysis was performed using the IB cluster method (Tishby, Pereira & Bialek, 2000) in order to compare item grouping (clustering) with the factors obtained in EFA. The aim of clustering is to place similar data in the same group to attain dimensionality reduction. Classical cluster algorithms have been applied in numerous subject areas including biology, marketing, artificial intelligence and motivational profile analysis (Gillet, Vallerand, & Rosnet, 2009; Ntoumanis et al., 2004; Vlachopoulos, Karageorghis, & Terry, 2000). While clustering can assist the analysis when large amounts of data are available, high levels of compression may give rise to information loss, hence a compromise between compression and relevance is necessary. The IB method, which is based on information theory, aims to compress data while retaining its relevance by seeking partitions of the variable *X* (clusters) that provide maximum information about the variable *Y*. The method has been applied successfully in various areas such as word-clusters for supervised and unsupervised text classification (Slonim & Tishby, 2001), galaxy spectra analysis (Slonim et al., 2001), image clustering (Goldberger, Greenspan, & Gordon, 2002) and speaker recognition (Hecht, Noor, & Tishby, 2009).

Given two variables *X* and *Y* with joint probability distribution *p(x,y)* and compression degree represented by variable *T*, the IB method attempts to minimize the function:

$$L[p(t|x)] = I(X,T) - \beta I(T,Y) \qquad (1)$$

where $I(X,T)$ is mutual information between the variables X and T, $I(Y,T)$ is mutual information between variables Y and T, and $\beta$ is the Lagrange multiplier representing the relationship between compression denoted by $I(X,T)$ and relevance denoted by $I(Y,T)$. When $\beta = 0$, compression is maximal and all relevant information is lost. However, relevance increases with $\beta$ and prevails when $\beta \to \infty$. For finite values of $\beta$, it is possible to compress X while maintaining a significant amount of information about Y. The IB method is characterized by the iterative algorithm comprising: (i) input of $p(x, y)$, $\beta$ and T, (ii) initialization of $p(t \mid x)$ with random values, and (iii) iteration according to the equations:

$$p(t_i) = \sum_j p(x_j) p(t_i \mid x_j) \qquad (2)$$

$$p(y_i \mid t_j) = \sum_k p(y_i \mid x_k) p(x_k \mid t_j) \qquad (3)$$

$$p(t_i \mid x_j) = \frac{p(t_i)}{Z(x,\beta)} EXP\left(-\beta \sum_k p(y_k \mid x_j) \log\left(p(y_k \mid x_j)/p(y_k \mid t_i)\right)\right) \qquad (4)$$

where Z is the partition function defined as:

$$Z(x,\beta) = \sum_i p(t_i) EXP\left(-\beta \sum_k p(y_k \mid x_j) \log\left(p(y_k \mid x_j)/p(y_k \mid t_i)\right)\right) \qquad (5)$$

After convergence, the algorithm finds probability distributions $p(t \mid x)$ and $p(y \mid t)$ that minimize the function given in equation (1). This distribution shows how variables are grouped, i.e. which variables belong to each cluster.

In the present study, joint distributions were drawn from a table in which x represented the responses of an individual to each item and y represented the items. For a fixed cluster number, the algorithm found the distribution of items per cluster in a manner that allowed clusters to be established according to how individuals had responded to them. The algorithm took into account only input data, i.e. items and responses of individuals, and it was not necessary to establish a similarity measure (as required by most other cluster algorithms) or a proportion between individual numbers and items (as in factor analysis).

EFA was carried out with the aid of the Statistical Program for the Social Sciences (SPSS), while cluster analysis was performed using the IB algorithm within the MatLab software package.

**RESULTS AND DISCUSSION**

The data collected using the 20 item instrument showed a KMO measure of sampling adequacy equal to 0.87 and the Bartlett's sphericity test gave a $\chi^2$ value of 1620.122 (*df = 190, p < 0,0*01), thus supporting the applicability of factor analysis (Hair et al., 2006). Cronbach's alpha coefficient was 0.87, which indicates a high level of reliability.

EFA yielded four eigenvalues greater than 1 that explained 57.76 % of data variance. As shown in Figure 2, the slope of the plot representing eigenvalues as a function of factor number decreased considerably from the fourth eigenvalue (elbow).

The factor matrix was subjected to varimax rotation in order to identify the factors, but some items presented cross loading. Following application of a second rotation using the quartimax method, we found four factors with eigenvalues greater than 1. The compositions of these factors and their respective Cronbach's alpha coefficients are presented in Table 1.

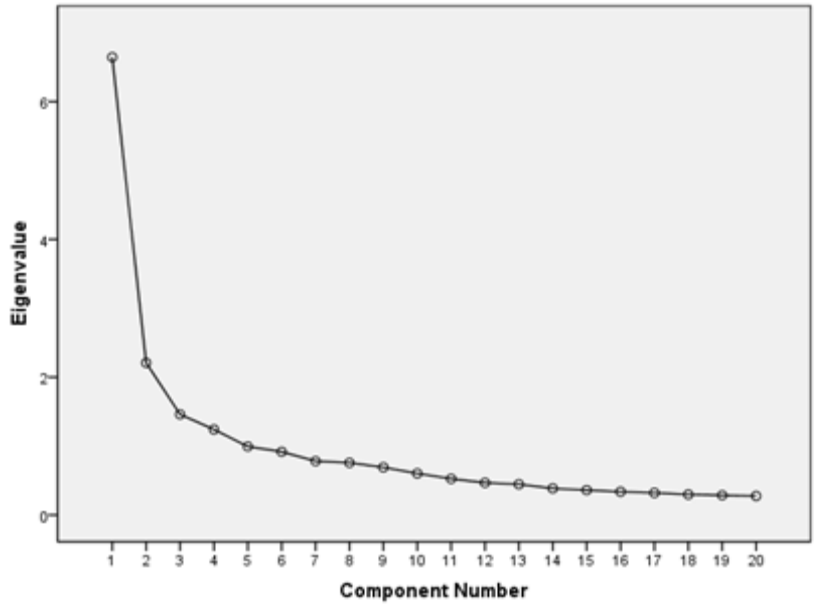

**Figure 2.** Eigenvalues plotted as a function of factor number.

**Table 1.** Factor Composition and Cronbach's Alpha Coefficients

| Factor composition | Items 1, 3, 4, 5, 6, 10, 13, 15, 17, 20 | Items 11, 12, 16, 19 | Items 9, 14, 18 | Items 2, 7, 8 |
|---|---|---|---|---|
| Cronbach's alpha coefficient | 0.87 | 0.74 | 0.76 | 0.42 |

Factor 4 presented a very low alpha coefficient indicating that items 2, 7 and 8 needed to be reviewed. While two of these items had a communality greater than 0.5, the communality of item 2 was very low (0.28) and so this item was removed and EFA was performed again with 19 items (Hair et al., 2006). The resulting factor structure was the same for the first three factors but factor 4, with only items 7 and 8, was difficult to interpret since the two items had different meaning. We suppose this occurred because there is a common word (family) in both items. In the analysis by Delgado (2008), item

7 did not load in any factor and, furthermore, the author found that, contrary to expectations, its meaning did not match with any factor. In addition, some authors consider that all factors should contain at least three items (O'Rourke & Hatcher, 2013).

By removing item 7 and performing EFA again, item 8 was reassigned to factor 1, resulting in a reduction in the total number of factors since only three remained with eigenvalues greater than one. Further analysis of factor structure employing both varimax and a quartimax rotation, revealed that all loadings per factor were greater than 0.4 (Table 2) and the recommendation of a sample size of at least 200 individuals was satisfied (Hair et al., 2006).

**Table 2.** Factor Loadings Associated with the Motivation for Visiting Questionnaire

| Item Number | Item Description | Factor Loading [a] |
|---|---|---|
| *Factor 1: Discovery News and Learning Motivation* | | |
| 1 | Discovery news | 0.75 |
| 3 | Having an opportunity to learn | 0.74 |
| 4 | Participating in the museum's activities | 0.54 |
| 5 | Entertainment | 0.59 |
| 6 | Satisfying my curiosity | 0.75 |
| 8 | Increasing my knowledge about science and that of my family/friends | 0.57 |
| 10 | Being challenged and having new experiences | 0.62 |
| 13 | Participating in museum activities to have an opportunity to interact with the experiments and to learn more | 0.59 |
| 15 | Satisfying my curiosity about science | 0.75 |
| 17 | Increasing my knowledge and that of my family/friends | 0.58 |
| 20 | Having an opportunity to learn science | 0.80 |
| *Factor 2: Interaction Motivation* | | |
| 11 | Feeling that people will see me in another way | 0.66 |

| 12 | Getting personal and professional skills | 0.56 |
| 16 | Being with people | 0.59 |
| 19 | Interacting with people | 0.71 |

*Factor 3: Leisure Motivation*

| 9 | Breaking the routine | 0.83 |
| 14 | Resting | 0.74 |
| 18 | "Killing" time | 0.76 |

Note: It was assumed that eigenvalues should be greater than 1 according to the Kaiser-Guttman criterion (Guttman, 1954; Kaiser, 1960).

As shown in Table 3, these factors explain 55.2% of data variance. An examination of the content of the three factors indicated that they were related to the five motivational components of the original scale that influence museum visiting. Factor 1, which contained 11 items, had acceptable internal consistency and reliability with a Cronbach's alpha coefficient of 0.844. This factor included the four *Discovery news motivation* items and three *Learning motivation* items together with items 4 and 13, which were derived from the *Participation in museum activities motivation,* and item 5, which was about entertainment but perceived by many visitors as a learning motivation. The relationship between entertainment/diversion and education/learning in informal settings has been discussed by a number of researchers (Falk, Moussouri, & Coulson, 1998; Packer, 2006).

Factor 1 was the most important of the three factors because it explained 35.18 % of the total of variation in the responses. We interpret this finding to mean that visitors considered that items related to discovery news, learning and participation in museum activities are so closely related that they grouped them as a set. This set represented one dimension by which visitors conceptualized their motivation for visiting the science museum. For simplicity, therefore, we labeled this factor Discovery news and Learning motivation.

**Table 3.** Cronbach's Alpha Coefficients, Percentage of Variance Explained, and Eigenvalues for Factors 1 to 3

| Factor | Cronbach's Alpha Coefficient | Percentage of Variance | Cumulative Percentage of Variance | Eigenvalue |
|---|---|---|---|---|
| 1 | 0.844 | 35.18 | 35.33 | 5.15 |
| 2 | 0.707 | 12.01 | 47.19 | 2.15 |
| 3 | 0.764 | 8.04 | 55.23 | 1.31 |

Notes: Factor 1 is *Discovery news and Learning motivation*; Factor 2 is *Interaction motivation*; Factor 3 is *Leisure motivation*.

These findings show that visitors perceived the science museum to be a setting that could best satisfy their educational interest needs. Numerous studies have concluded that learning motivation, as a predominant factor, plays a decisive role in influencing museum visitation (Falk & Dierking, 2000; Ji et al., 2014; Packer, 2006). In contrast, low-income families in Taiwan are apparently deterred from visiting museums because of their strong association with education. According to Lin (2006), this segment of the population would prefer/seek museums as environments for exploration and entertainment as well as for learning.

In order to incorporate learning opportunities into an exhibition, museums need to develop specific activities that give visitors both the possibility to expand their understanding of particular science subjects and the chance to relate science and technology with their everyday life. Activities that allow controlled choice are more suitable because they encourage effective and more complex learning (Bamberger & Tal, 2007; Griffin, 2004). In addition, museum programs can stimulate teachers in the planning and management of school excursions by, for example, linking the topics being studied at school with those on display at the museum (Griffin & Symington, 1997).

Factor 2 was the second most important of the three by explaining 12.01 % of the total variation in the responses. Since the three items in this factor comprised three of the four interaction motivational components, the label *Interaction motivation* was retained. Interestingly, item 12 was loaded in factor 2 even though this item is related to learning and was expected to be assigned to factor 1.

The last of the three factors accounted for 8.04 % of the total variation in responses. Since this factor contained the remaining three of the five leisure and entertainment-motivational components, it was given the label *Leisure motivation*.

According to exploratory factor analysis of the responses, three factors were perceived, namely learning, interaction and leisure. As a comparative analysis, we investigated the clustering of the final 18 items included in the EFA by applying the IB method since it does not depend on individual numbers. Initial tests were performed to determine a $\beta$ value for each parameter $T$, following which joint probabilities $p\ (x,y)$ between items and individuals responses were calculated and the IB algorithm applied to find solutions that minimized the function given in equation (1). All of the data was compressed into one cluster for $T = 1$, while two clusters ($T_1$ and $T_2$) were obtained for $T = 2$ as detailed in Figure 3. It is noteworthy that cluster $T_2$ contained the same items as EFA factor 3 (*Leisure motivation*), namely 9, 14 and 18. When the IB method was applied with $T = 3$, cluster $T_2$ remained unchanged but cluster $T_1$ was split into two, the second of which contained the same items as EFA factor 2 (*Interaction motivation*) with the single omission of item 12, which was placed in the first cluster together with the 11 items constituting EFA factor 1 (*Discovery news and Learning motivation*). Since the only difference between factor and cluster structures related to item 12, this item was omitted from the response data and EFA and the IB method were performed once more. As shown in Figure 4, the results coincided with both factors and clusters showing the same item structure.

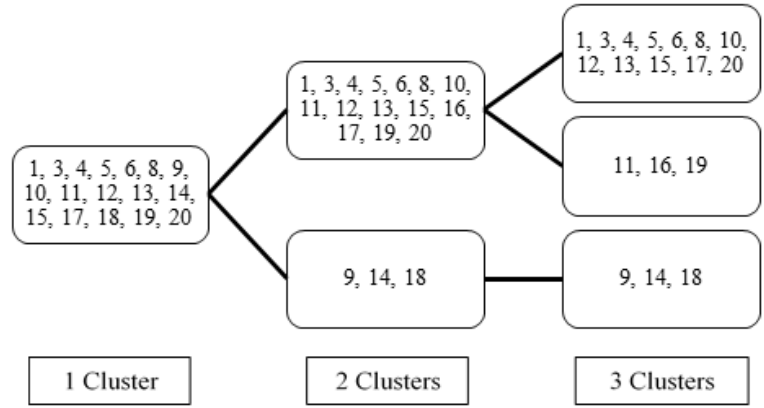

**Figure 3.** Item clusters formed using the IB method.

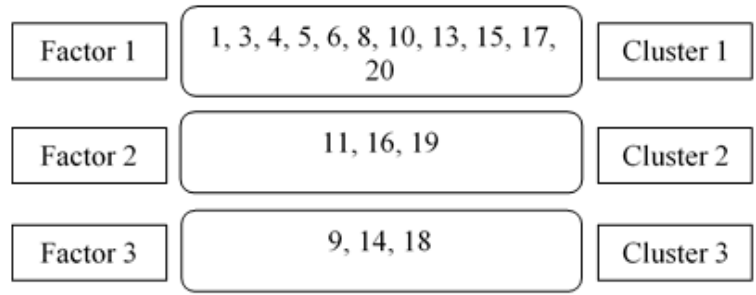

**Figure 4.** Factors and clusters formed after removing item 12.

Considering only the final 17 items, we analyzed the scores by gender but no significant differences could be found between males and females with regard to total scores on the motivation for visiting questionnaire. In order to assess age-related differences in motivation, respondents were divided into six age groups but, as for gender, the analyses indicated that respondents in the various age groups did not differ significantly in motivation.

**CONCLUSIONS**

The instrument elaborated in the present study, although originally developed for the "Pavilion of Knowledge" in Lisbon, Portugal, can be applied to assess the quality of other heritage sites because the methodology is flexible and can be adjusted to specific tourist entities and their specific contexts. As the questionnaire was originally developed in Portugal, it had to be evaluated and modified to be of use in the Brazilian context. Application of the instrument in other cultures and languages would likely require further verification of its validity and reliability.

The evaluation of the instrument in the Science Museum "Professor Mário Tolentino" revealed that EFA and IB can be used as complementary methods to modify the tool and provide insights into visitor motivation. The survey showed that motivations to visit the institute were similar in male and female respondents of all age groups. Seeking opportunities to learn and to interact with other persons are two significant motivations to visit, and these have to be satisfied by the science museum studied. Such studies are under-represented in the Brazilian perspective and the results of this type of survey enrich our understanding of the cultural diversity of museum contexts.

**FUTURE RESEARCH**

We hope that, after minor modification, the instrument described herein can be analyzed by confirmatory factor analysis with a large sample size and will ultimately be useful for application in other informal environments in order to promote debate about visitor motivations in Brazil. Alongside research possibilities, we see merit in further exploration of the visiting strategies that have been employed by the studied museum since the findings may suggest changes that could have key planning, management, and marketing implications for the future implementation of a Chemistry Gallery.

Another important activity for future research in this area would be to consider the findings of our final survey in combination with results obtained using qualitative methods, such as interviews and essays, with the aim of obtaining deeper insights into the motivation for visiting the institute. According to Parasuraman, Zeithaml and Berry (1988), application of an instrument of the type described herein acquires particular value when it is used periodically, and when it is employed in conjunction with other forms of product quality measurement.

**REFERENCES**

Associação Brasileira de Centros e Museus de Ciência. **Centros e museus de ciência do Brasil 2015**. Rio de Janeiro: UFRJ. FCC. Casa da Ciência; Fiocruz. Museu da Vida, 2015. Retrieved from http://www.mcti.gov.br/documents/10179/472850/Centros+e+Museus+de+Ci%C3%AAncia+do+Brasil+2015+-+pdf/667a12b2-b8c0-4a37-98f5-1cbf51575e63. Accessed April 10, 2016.

Bamberger, Y., & Tal, T. (2007). Learning in a personal context: Levels of choice in a free choice learning environment in science and natural history museums. *Science Education, 91*(1), 75-95.

Bjørnebekk, G. (2009). Psychometric properties of the scores on the behavioral inhibition and activation scales in a sample of Norwegian children. *Educational and Psychological Measurement*, *69*(4), 636-654.

Costello, A. B., & Osborne, J. W. (2005). Best practices in exploratory factor analysis: four recommendations for getting the most from your analysis. *Practical Assessment, Research & Evaluation, 10*(7), 1-9.

Delgado, M. F. S. (2008). *Constrangimentos às visitas aos centros de ciência: o caso do Pavilhão do Conhecimento* (Masters dissertation, Universidade de Aveiro, Aveiro, Portugal). Retrieved from http://hdl.handle.net/10773/1623.

Falk, J. H. (2006). An identity-centered approach to understanding museum learning. *Curator: The Museum Journal, 49*(2), 151–166.

Falk, J. H., & Dierking, L. D. (1992). *The museum experience*. Washington, DC: Whalesback Books.

Falk, J. H., & Dierking, L. D. (2000). *Learning from museums:* Visitor experiences and the making of meaning. Lanham, MD: AltaMira Press.

Falk, J. H., Moussouri, T., & Coulson, D. (1998). The effect of visitors' agendas on museum learning. *Curator: The Museum Journal, 41*(2), 107-120.

Gillet, N., Vallerand, R. J., & Rosnet, E. (2009). Motivational clusters and performance in a real-life setting. *Motivation and Emotion, 33*(1), 49-62.

Glynn, S. M., Taasoobshirazi, G., & Brickman, P. (2009). Science Motivation Questionnaire: construct validation with nonscience majors. *Journal of Research in Science Teaching, 46*(2), 127–146.

Goldberger, J., Greenspan, H., & Gordon, S. (2002). Unsupervised image clustering using the information bottleneck method. In L. Van Gool (Ed.), *Pattern recognition* (pp. 158-165). Heidelberg, Germany: Springer.

Griffin, J. (2004). Research on students and museums: Looking more closely at the students in school groups. *Science Education, 88*(Suppl.1), S59-S70.

Griffin, J., & Symington, D. (1997). Moving from task-oriented to learning-oriented strategies on school excursions to museums. *Science Education, 81*(6), 763-779.

Guttman, L. (1954). Some necessary conditions for common-factor analysis. *Psychometrika, 19*(2), 149-161.

Hair, J. F., Black, B., Babin, B., Anderson, R. E., & Tatham, R. L. (2006). *Multivariate data analysis* (6th ed.). Upper Saddle River, NJ: Pearson Education.

Hecht, R. M., Noor, E., & Tishby, N. (2009). Speaker recognition by Gaussian information bottleneck. In *Proceedings of the 10th Annual Conference of the International Speech Communication Association 2009* (INTERSPEECH 2009, September, Brighton, UK, pp. 1567-1570).

Ji, J., Anderson, D., Wu, X., & Kang, C. (2014). Chinese family groups' museum visit motivations: A comparative study of Beijing and Vancouver. *Curator: The Museum Journal, 57*(1), 81-96.

Kaiser, H. F. (1960). The application of electronic computers to factor analysis. *Educational and Psychological Measurement, 20*(1), 141-151.

Lin, Y. (2006). Leisure - A function of museums? The Taiwan perspective. *Museum Management and Curatorship, 21*(4), 302-316.

Massidda, D., Carta, M. G., & Altoè, G. (2016). Integrating Different Factorial Solutions of a Psychometric Tool Via Social Network Analysis. *Methodology*, *12*(3), 97-106.

Ntoumanis, N., Pensgaard, A. M., Martin, C., & Pipe, K. (2004). An idiographic analysis of amotivation in compulsory school physical education. *Journal of Sport and Exercise Psychology, 26*(2), 197-214.

O'Rourke, N., & Hatcher, L. (2013). *A step-by-step approach to using SAS for factor analysis and structural equation modeling* (2nd ed.). Cary, NC: SAS Institute.

Packer, J. (2006). Learning for fun: The unique contribution of educational leisure experiences. *Curator: The Museum Journal, 49*(3), 329-344.

Packer, J., & Ballantyne, R. (2002). Motivational factors and the visitor experience: A comparison of three sites. *Curator: The Museum Journal, 45*(3), 183-198.

Parasuraman, A., Zeithaml, V. A., & Berry, L. L. (1988). Servqual: A multiple-item scale for measuring consumer perceptions of service quality. *Journal of Retailing, 64*(1), 12-40.

Robinson, J. P., Shaver, P. R., & Wrightsman, L. S. (1991). Criteria for scale selection and evaluation. In *J. P. Robinson, P. R. Shaver & L. S. Wrightsman (Eds.), Measures of personality and social psychological attitudes, 1*(3), pp.1-16. Amsterdam: Elsevier.

Slonim, N., & Tishby, N. (2001). The power of word clusters for text classification. In *Proceedings of the 23rd European Colloquium on Information Retrieval Research* (ECIR 2001, Darmstadt, Germany, 200-212).

Slonim, N., Somerville, R., Tishby, N., & Lahav, O. (2001). Objective classification of galaxy spectra using the information bottleneck method. *Monthly Notices of the Royal Astronomical Society, 323*(2), 270-284.

Studart, D. C. (2005). Museus e famílias: percepções e comportamentos de crianças e seus familiares em exposições para o público infantil. *História, Ciências, Saúde – Manguinhos, 12*(suppl), 55-77.

Tabachnick, B. G., & Fidell, L. S. (2007). *Using multivariate statistics* (5th ed.). Boston, MA: Pearson/Allyn & Bacon.

Tishby, N., Pereira, F. C., & Bialek, W. (2000). The information bottleneck method. arXiv preprint physics/0004057.

Tze, V. M., Klassen, R. M., Daniels, L. M., Li, J. C. H., & Zhang, X. (2013). A cross-cultural validation of the Learning-Related Boredom Scale (LRBS) with Canadian and Chinese college students. *Journal of Psychoeducational Assessment*, *31*(1) 29-40.

Vlachopoulos, S. P., Karageorghis, C. I., & Terry, P. C. (2000). Motivation profiles in sport: A self-determination theory perspective. *Research Quarterly for Exercise and Sport, 71*(4), 387-397.

Williams, B., Onsman, A., & Brown, T. (2010). Exploratory factor analysis: a five-step guide for novices. *Journal of Emergency Primary Health Care, 8*(3), Article 990399.

Zwinkels, J., Oudegeest, T., & Laterveer, M. (2009). Using visitor observation to evaluate exhibits at the Rotterdam Zoo Aquarium. *Visitor Studies, 12*(1), 65-77.

# APPENDIX 1

## The Twenty Items Initially Employed to Assess the Motivation for Visiting

1 - Discovery News **(2)**

2 - Making myself comfortable and under no pressure **(1)**

3 - Having an opportunity to learn **(3)**

4 - Participating in museum's activities **(5)**

5 - Entertainment **(1)**

6 - Satisfying my curiosity **(2)**

7 - Going with children/family members who are interested in the museum's activities **(4)**

8 - Increasing my knowledge and that of my family/friends about science **(3)**

9 - Breaking the routine **(1)**

10 - Being challenged and having new experiences **(2)**

11 - Feeling that people will see me in another way **(4)**

12 - Getting personal and professional skills **(3)**

13 - Participating in museum's activities to have an opportunity to interact with the experiments and to learn more **(5)**

14 - Resting **(1)**

15 - Satisfying my curiosity about science **(2)**

16 - Being with people **(4)**

17 - Increasing my knowledge and that of my family/friends **(3)**

18 - "Killing" time **(1)**

19 - Interacting with people **(4)**

20 - Having an opportunity to learn science **(3)**

Notes: Adapted from Delgado (2008)

The proposed factor of each item is shown in parentheses according to the following:

**(1)** Leisure and entertainment motivation

**(2)** Discovery news motivation

**(3)** Learning motivation

**(4)** Interaction motivation

**(5)** Participation in museum activities motivation